\advance\voffset by 1mm
\advance\hoffset by -7mm
\makeatletter
\@addtoreset{equation}{section}
\makeatother

\renewcommand{\title}[1]{\null\vspace{25mm}

\noindent{\Large{\bf #1}}\vspace{10mm}  

\noindent {\large By }}
\newcommand{\authors}[1]{\noindent{\large #1}\vspace{3mm}

}
\newcommand{\address}[1]{\noindent #1\vspace{5mm}

}

\documentstyle[11pt,epsfig]{article}

\newtheorem{th}{Theorem} 
\newtheorem{lm}{Lemma}
\newtheorem{df}{Definition}  
\newtheorem{pr}{Proposition} 
\newtheorem{as}{Assumption}
\newcommand{\bth}{\begin{th}}
\newcommand{\eth}{\end{th}}  
\newcommand{\blm}{\begin{lm}} 
\newcommand{\elm}{\end{lm}} 
\newcommand{\bdf}{\begin{df}} 
\newcommand{\edf}{\end{df}} 
\newcommand{\bpr}{\begin{pr}}
\newcommand{\epr}{\end{pr}}
\newcommand{\bas}{\begin{as}}
\newcommand{\eas}{\end{as}}
\newcommand{\bpf}{\noindent {\bf Proof:} }
\newcommand{\epf}{$\bullet$\par\vspace{1.8mm}\noindent}
\newcommand{\beq}{\begin{equation}} 
\newcommand{\eeq}{\end{equation}\par\noindent}
\newcommand{\bit}{\begin{itemize}}
\newcommand{\eit}{\end{itemize}\par\noindent}
\newcommand{\beqa}{\begin{eqnarray*}}
\newcommand{\eeqa}{\end{eqnarray*}\par\noindent}
\newcommand{\beqn}{\begin{eqnarray}}
\newcommand{\eeqn}{\end{eqnarray}\par\noindent}
\newcommand{\la}{\lambda} 
\newcommand{\La}{\Lambda}
\newcommand{\si}{\sigma} 
\newcommand{\Si}{\Sigma}
\newcommand{\DL}{\Delta\Lambda}

\begin{document}

\newfont{\sdub}{msym7} 
\newfont{\dub}{msym7 scaled\magstep1}
\newfont{\ldub}{msym9 scaled\magstep1}

\newfam\msbfam
\font\tenmsb=msbm10			\textfont\msbfam=\tenmsb
\font\sevenmsb=msbm7		\scriptfont\msbfam=\sevenmsb
\font\fivemsb=msbm5			\scriptscriptfont\msbfam=\fivemsb
\def\Bbb{\fam\msbfam \tenmsb}

\def\insim{\, \raise-5 pt\hbox{$\buildrel\hbox{\hbox{$\sim$ \hskip - 12 truept \raise
5pt\hbox{$\scriptstyle \in$}}}\over{\scriptstyle\,\,}\ $}}
\def\duni{\raise-3 pt\hbox{$\buildrel\hbox{\hbox{$\bigcirc$ \hskip - 13 truept
\raise -0.1pt\hbox{$\bf \cup$}}}\over{\scriptstyle\,\,} $}\, }
\def\dunia{\raise-7 pt\hbox{$\buildrel\hbox{\hbox{$\bigcirc$ \hskip - 13.0 truept 
\raise -0.1pt\hbox{$\bf \cup$}\hskip - 6.0 truept\raise-6.5pt\hbox{$\scriptstyle
a$}}}\over{\scriptstyle\,\,} $}\,  \hskip 1.5pt}
\def\dsuma{\raise-7 pt\hbox{$\buildrel\hbox{\hbox{$\bigcirc$ \hskip - 13.7 truept
\raise -0pt\hbox{$\bf +$}\hskip - 6.5 truept\raise-6.5pt\hbox{$\scriptstyle
a$}}}\over{\scriptstyle\,\,} $}\, \hskip 1pt}
\def\sdsuma{\raise-5.1 pt\hbox{$\buildrel\hbox{\hbox{$\scriptstyle\bigcirc$ \hskip -
11.2 truept \raise -0pt\hbox{$\scriptstyle\bf +$}\hskip - 5.3
truept\raise-4.8pt\hbox{$\scriptscriptstyle a$}}}\over{\scriptscriptstyle\,} $}\,
\hskip 0.7pt}
\def\nonexist{\raise-8 pt\hbox{$\buildrel\hbox{\hbox{$\exists$ \hskip - 9.0 truept 
\raise -0.1pt\hbox{$|$}\hskip - 6.0 truept\raise-6.5pt\hbox}}\over{\scriptstyle\,\,} $}\hskip 8pt}
\def\subsim{\, \raise-5 pt\hbox{$\buildrel\hbox{\hbox{$\sim$ \hskip - 13 truept \raise
5pt\hbox{$\scriptstyle \subset$}}}\over{\scriptstyle\,\,}\ $}}
\def\eqsim{\ \raise-3.5 pt\hbox{$\buildrel\hbox{\hbox{$\sim$ \hskip - 12 truept \raise
3.5pt\hbox{$\scriptstyle =$}}}\over{\scriptstyle\,\,}\ $}\hskip +1 truept}
 
\title{A Classification of Classical Representations for Quantum-like Systems.}  
\authors{Bob Coecke} 
\address{FUND-DWIS, Free University of Brussels,\\ Pleinlaan 2,
B-1050 Brussels, Belgium;\\ bocoecke@vub.ac.be}
\par
\vspace{-7.8cm}
\par
\noindent
This paper has been published in {\it Helv. Phys. Acta} {\bf 70}, 462 (1997).
\par
\vspace{8.8cm}
\par
\noindent
For general non-classical systems, we study the different classical representations that fulfill
the specific context dependence imposed by the hidden measurement system formalism introduced in
\cite{axiomatics}.  We show that the collection of non-equivalent representations has a poset
structure. We also show that in general, there exists no 'smallest' representation, since this poset
is not a semi-lattice. Then we study the possible representations of quantum-like measurement
systems. For example, we show that there exists a classical representation of finite dimensional
quantum mechanics with ${\Bbb N}$ as a set of states for the measurement context, and we build
an explicit example of such a representation.  

\section{Introduction.}   
   
For a detailed introduction on the general subject we refer
to \cite{axiomatics}, where we have proved that every 'measurement system'
(=m.s.) has a representation as a 'hidden measurement system' (=h.m.s.), i.e., every physical entity
that is defined by a collection of states and a collection of measurements, such that for a
defined initial state we have a probability measure that defines the relative occurrence of the
outcomes, there exist a classical representation with a very specific kind of context
dependence.  Thus, also for quantum entities there exist such a classical representation.
An obvious question is: which different classical representations exist for a given
quantum entity.  In this paper we build a classification of all possible
h.m.s.-representations, for every given quantum m.s.
A lot of preparing work has already be done in \cite{axiomatics}, in section 4 and in the appendix.
Therefore, we will regularly refer to the theorems and definitions of that paper.
For the identification of the h.m.s.-representations for quantum m.s$.$ we proceed in three steps. 
First we show that the collection of possible h.m.s.-representations has a poset structure, and this
will allow us to characterize the possible h.m.s.-representations for a given m.s$.$ by its smallest
h.m.s.-representations.  Then we show that in general, there exists no 'preferred' smallest
h.m.s.-representation, and this forces us to consider the quantum m.s$.$ case by case. After these
three steps we present two examples of explicit h.m.s.-representations which illustrate the
results that we have obtained in this paper.  As in \cite{axiomatics}, for a general definition of
the basic mathematical objects we refer to
\cite{bir40} and \cite{sik69}. As in \cite{axiomatics}, we remark that the
results presented in this paper
where made known in \cite{thesisbob}.

\section{An additional assumption.}

In this paper we make a from a physical point of view very natural assumption, namely we
only consider h.m.s$.$ for which there exists a probability measure $\mu\in {\bf MX}$ such 
that
this h.m.s$.$ belongs to ${\bf HMS}(\mu)$, i.e., for every measurement of this h.m.s$.$ we
consider the same set of states of the measurement context and the same 
relative frequency of
occurrence of these states.  The motive of this additional assumption is essentially
'simplicity', since a general treatment (which is indeed possible) would not lead to any
new conceptual insights or mathematical insights, but would only make things a lot more
complicated.
Nonetheless, in the light of section 3.3 of \cite{axiomatics} this assumption is also very
natural
since most m.s$.$ that we consider in practice have a sufficient additional
structure to impose a representation for all measurements, starting with a representation
for one measurement, and all of the with the same probability measure that determines
the relative frequency of occurrence of the states of the measurement context.
We also remark that this assumption does not influence the results of
\cite{axiomatics}, and in particular does the existence theorem in section 4.3 remain valid.
For this subset of ${\bf HMS}$ that we consider in this paper we introduce the following
notation:
\beq
{\bf HMS}({\bf MX})=\{\Si,{\cal E}|\Si,{\cal E}\in{\bf HMS}(\mu),\mu\in {\bf MX}\}  
\eeq

\section{A poset characteristic for possible representations.}

In section 4.1 in \cite{axiomatics} we have defined a set ${\bf M}$, consisting of classes of
isomorphic measure spaces.  We also proved that for every separable measure space, there exists
one and only one class in ${\bf M}$ to which this measure space belongs (see Lemma 1 of
\cite{axiomatics}).  In Definition 8 of \cite{axiomatics} we introduced a binary relation $\leq$ on
${\bf M}$.  We have the following lemma:
\blm
${\bf M}, \leq$ is a poset.
\elm  
\bpf
The proof that $\leq$ is reflexive and transitive is straightforward, so we only have to
verify whether it's anti-symmetrical.  Let ${\cal M}\leq {\cal M}'$ and ${\cal M}'\leq
{\cal M}$, and ${\cal B},\mu\in {\cal M}$, ${\cal B}',\mu'\in {\cal M}$, 
and $F: {\cal B}\rightarrow {\cal B}'$ and
$F': {\cal B}'\rightarrow {\cal B}$ fulfilling Definition 8 in \cite{axiomatics}. First we
prove that, given the above stated assumptions (the existence of $F$ and $F'$), there are two
possibilities: ${\cal M}, {\cal M}'\in 
{\bf M}_{{\Bbb X}}\cup {\bf M}_{{\Bbb N}}$ and 
${\cal M}, {\cal M}'\in {\bf M}_{{\Bbb R},a}$. 
Suppose that ${\cal M}\in {\bf M}_{{\Bbb R},a}$ and ${\cal M}'\in {\bf M}_{{\Bbb
R},a'}$ with 
$a\not=a'$. Consider the sets $B_0=(\emptyset, I)\in {\cal B}$,
and $B'_0=(\emptyset, I')\in {\cal B}'$ ($I$ and $I'$ are the greatest elements of different Borel
algebras). We have $F(B_0)\not=B'_0$ since $\mu'(F(B_0))=\mu(B_0)=a$ and
$\mu'(B'_0)=a'\not=a$. Thus, $F(B_0^c)\cap B'_0\not=\emptyset$ or $F(B_0^c)\subset
{B'}_0^c$.  Suppose $F(B_0^c)\cap {B'}_0\not=\emptyset$. According to Lemma 3 in
\cite{axiomatics},
${\cal B}_l,\mu_l$ is a measure space (we also use the notations of Lemma 3 in
\cite{axiomatics}), and, following Lemma 2 in \cite{axiomatics} we find
${\cal B}_l,\mu_l\cong {\cal B}_{{\Bbb R}},\mu$. As a consequence, ${\cal B}_{{\Bbb
R}}\cong{\cal B}_l\cong\{F(B)|B\in {\cal B}_l\}$ (due to Proposition 3 in \cite{axiomatics},
$\{F(B)|B\in {\cal B}_l\}$ is a Borel subalgebra of ${\cal B}'$, isomorphic with ${\cal
B}_l$). But, since $F(B_0^c)\cap B'_0\not=\emptyset$, this situation contradicts with
the fact that ${\cal B}_{{\Bbb R}}$
cannot be embedded in a one-to-one way in ${\cal B}_{{\Bbb N}}$. If
$F(B_0^c)\subset {B'}_0^c$ we can make the same reasoning by exchanging the roles of
${\cal B},\mu$ and  
${\cal B}',\mu'$. In an analogous way one proves that the following two situations:
${\cal M}\in {\bf M}_{{\Bbb R}}$, ${\cal M}'\in {\bf M}_{{\Bbb X}}\cup {\bf
M}_{{\Bbb N}}$ and ${\cal M}\in {\bf M}\setminus \{{\cal M}_{{\Bbb R}}\}$,
${\cal M}'={\cal M}_{{\Bbb R}}$ are not possible.   
Suppose that ${\cal M}, {\cal M}'\in {\bf M}_{{\Bbb N}}$. Let ${\cal
B}={\cal B}'={\cal B}_{{\Bbb N}}$, $\mu=\mu_m$ and $\mu'=\mu_{m'}$. If
$i,j\in{\Bbb N}$ and $i\not=j$ then $F(\{i\})\cap F(\{j\})=F(\{i\}\cap
\{j\})=F(\emptyset)=\emptyset$. Let $n_-$ be the smallest integer such that $F(\{n_-\})$ 
is not a singleton,
and let
$n_+$ be the largest integer such that $m(n_+)=m(n_-)$. Since $(m(i))_i$ is a decreasing sequence, 
$\bigl(\mu_{m'}(F(\{i\}))\bigr)_i=\bigl(\mu_{m}(\{i\})\bigr)_i=(m(i))_i$ is also a
decreasing sequence. Since we have that: 
\beqa 
&1) &\bigl(\mu_{m'}(F(\{i\}))\bigr)_i {\rm \ is\ a\ decreasing\ sequence}\\ 
&2) &\bigl(\mu_{m'}(i)\bigr)_i=(m'(i))_i {\rm \ is\ a\ decreasing\ sequence}\\ 
&3) &\forall i\in{\Bbb N},\exists j\in{\Bbb N}{\rm \ such\ that\ } i\in F(\{j\})\\
&4) &\forall i\in{\Bbb X}_{n_--1}:F(\{i\}) {\rm \ is\ a\ singleton} 
\eeqa
we have $\cup_{j=1}^{j=n_--1}F(\{j\})={\Bbb
X}_{n_--1}$. Moreover, since for all $i\in F(\{n_-\}):\mu_{m'}(F(\{n_-\})\setminus
\{i\})>0$, we can conclude with the following image:
\beqa
\forall i\in\cup_{j=1}^{j=n_--1}F(\{j\})&:&
m'(i)=\mu_{m'}(\{i\})=\mu_{m'}(F(\{i\}))=m(i)\geq m(n_-)\\  
\forall i\in F(\{n_-\})&:& 
m'(i)=\mu_{m'}(\{i\})<\mu_{m'}(F(\{n_-\}))=m(n_-)\\ 
\forall i\in\cup_{j=n_-+1}^{j=n_+}F(\{j\})&:& 
m'(i)\leq m(n_-)\\ 
\forall i\not\in\cup_{j=1}^{j=n_+}F(\{j\})&:& 
m'(i)<m(n_-)
\eeqa  
In the case that $n_-+1\leq j\leq n_+$, $i\in F(\{j\})$ and 
$m'(i)= m(n_-)$, we have $\{i\}=F(\{j\})$. As a consequence, there are
at most
$n_+-1$ integers
$i$ such that
$m'(i)\geq m(n_+)=m(n_-)$, and thus we have $m(n_+)>m'(n_+)$.
All this results in the following equations:
\beq\label{m(i)=m'(i), m(n)>m'(n)}
\forall i\in{\Bbb X}_{n_+-1}: m(i)\geq m'(i), m(n_+)>m'(n_+)
\eeq 
Analogously, with the same reasoning on $F'$, we find that there exists an integer $n_+'$
such that:
\beq\label{m(i)=m'(i), m(n')<m'(n')}
\forall i\in{\Bbb X}_{n_+'-1}: m(i)\leq m'(i), m(n_+')<m'(n_+')
\eeq
The contradiction between eq.\ref{m(i)=m'(i), m(n)>m'(n)} and 
eq.\ref{m(i)=m'(i), m(n')<m'(n')} indicates that there exist no such integers $n_+$ and
$n'_+$. As a consequence there exist no integers $n_-$ and
$n'_-$, and thus, $F$ and $F'$ are onto, i.e., they are isomorphisms of measure spaces (see
Proposition
3 in \cite{axiomatics}). As a consequence, ${\cal B},\mu\cong {\cal B}',\mu'$, 
and thus we have ${\cal M}={\cal M}'$. 
The case ${\cal M}, {\cal M}'\in{\bf M}_{{\Bbb X}}$ proceeds along the same
lines. It is also clear that the case  ${\cal M}\in {\bf M}_{{\Bbb X}}$ and ${\cal
M}'\in {\bf M}_{{\Bbb N}}$ is not possible. For the case  ${\cal M}, {\cal
M}'\in {\bf M}_{{\Bbb R},a}$, we already know from the foregoing part of this proof that
$F(B_0)=B'_0$. Along the same lines as for the case ${\cal M}, {\cal M}'\in
{\bf M}_{{\Bbb X}}\cup {\bf M}_{{\Bbb N}}$, and by applying Lemma 3 in
\cite{axiomatics}, we find that ${\cal M}={\cal M}'$.   
\epf
Moreover, ${\bf M},\leq$ has a greatest element, namely ${\cal M}_{{\Bbb R}}$ (see Lemma
6 in
\cite{axiomatics}).  Thus, ${\bf M},\leq$ is a poset with a greatest element.  As we will show now,
this poset characterizes the possible different candidates for h.m.s.-representations.
First we introduce 'inclusion up to mathematical equivalence' based on the definition of
'belonging up to mathematical equivalence' (see \cite{axiomatics}).
If $\Sigma,{\cal E}\in {\bf N}$ implies that $\Sigma,{\cal E}\insim {\bf N}$ we
write: 
\beq
{\bf N}\subsim {\bf N}'
\eeq
If both ${\bf N}\subsim {\bf N}'$ and ${\bf N}'\subsim {\bf N}$ are valid, we
write:
\beq
{\bf N}\eqsim {\bf N}'
\eeq
One easily sees that the condition '$\Sigma,{\cal E}\in {\bf N}$ implies that
$\Sigma,{\cal E}\insim {\bf N'}\,$' is equivalent with '$\Sigma,{\cal E}\insim {\bf N}$
implies that $\Sigma,{\cal E}\insim {\bf N'}\,$'. Thus, if ${\bf N}\eqsim {\bf N}'$, then
${\bf N}$ and ${\bf N}'$ are equivalent sets of representations, i.e$.$, if a measure
space has a representation $\Si,{\cal E}$ in ${\bf N}$, then it has a representation
$\Si',{\cal E}'$ in ${\bf N}'$ such that $\Si,{\cal E}\sim \Si',{\cal E}'$, and vice
versa.  As we saw in section 4.2 in \cite{axiomatics}, with every $\mu\in{\bf MX}$  we can relate
${\cal M}_\mu\in{\bf M}$.  For every ${\cal M}\in{\bf M}$ we can introduce the collection of all
h.m.s$.$ in ${\bf HMS}({\bf MX})$ which are such that ${\cal M}_\mu={\cal M}$:
\bit
\item ${\bf HMS}({\cal M})=\cup_{{\cal M}_\mu={\cal M}}{\bf HMS}(\mu)$
\eit
This allows us to introduce the following collection of classes of h.m.s$.$:
\bit
\item ${\bf M}_{\bf HMS}=\{{\bf HMS}({\cal M})|{\cal M}\in{\bf M}\}$
\eit
The importance if this collection of classes of h.m.s$.$ follows from the following
two theorems:
\bth\label{th:equicolrep}
$\forall\mu,\mu'\in{\bf MX}: {\cal M}_\mu ={\cal M}_{\mu'}\Rightarrow{\bf
HMS}(\mu)\eqsim{\bf HMS}(\mu')$.
\eth
\bpf
If $\Si,{\cal E}\in{\bf HMS}(\mu)$, then $\Si,{\cal E}\insim{\bf
HMS}(\mu)$. 
As a consequence of Theorem 1 in \cite{axiomatics} we have $\Delta{\bf
M}(\Si,{\cal E})\leq {\cal M}_\mu={\cal M}_{\mu'}$,
and thus, again due to
Theorem 1 in \cite{axiomatics}, $\Si,{\cal E}\insim{\bf HMS}(\mu')$. This results in ${\bf
HMS}(\mu)\subsim{\bf HMS}(\mu')$. Analogously, we find ${\bf
HMS}(\mu')\subsim{\bf HMS}(\mu)$, and thus ${\bf
HMS}(\mu)\eqsim{\bf HMS}(\mu')$.
\epf
Thus, for a given m.s., if $\mu$ and $\mu'$ are such that ${\cal
M}_\mu={\cal M}_{\mu'}$, then ${\bf HMS(\mu)}$ and ${\bf HMS(\mu')}$ are equivalent
collections of h.m.s.-representations. As a consequence, for all ${\cal M}\in
{\bf M}$, ${\bf HMS}({\cal M})$ can be considered as a
collection of equivalent h.m.s$.$-representations.
The different collections of equivalent h.m.s$.$-representations are ordered in the
following way.
\bth\label{th:subifleqiflsim}
${\bf M},\leq$ and ${\bf M}_{\bf HMS}, \subsim$ are isomorphic posets.
\eth
\bpf
We have to prove that 
${\bf HMS}({\cal M})\subsim {\bf HMS}({\cal M}')\Leftrightarrow {\cal M}\leq {\cal M}'$.
\par\noindent i) ${\cal M}\leq {\cal M}'\Rightarrow {\bf HMS}({\cal M})\subsim {\bf HMS}({\cal M}')$:
Since $\Si,{\cal E}\in{\bf HMS}({\cal M})$ implies $\Si,{\cal E}\insim{\bf HMS}({\cal
M})$, there exists $\mu\in {\bf MX}$ such that $\Si,{\cal E}\insim{\bf HMS}(\mu)$.  As a
consequence of Theorem 1 in \cite{axiomatics} we have $\Delta{\bf M}(\Si,{\cal
E})\leq {\cal M}_\mu={\cal M}$. If ${\cal M}\leq {\cal M}'$ then   $\Delta{\bf M}(\Si,{\cal
E})\leq {\cal M}'={\cal M}_{\mu'}$ for every $\mu'$  such that ${\cal M}_{\mu'}={\cal M}'$.
Thus, again due to Theorem 1 in \cite{axiomatics} we have $\Si,{\cal E}\insim{\bf
HMS}(\mu')\subsim{\bf HMS}({\cal M}')$, and thus ${\bf HMS}({\cal M})\subsim {\bf HMS}({\cal
M}')$. 
\par\noindent ii) ${\bf HMS}({\cal M})\subsim {\bf HMS}({\cal M}')\Rightarrow {\cal M}\leq {\cal M}'$:
If ${\bf HMS}({\cal M})\subsim {\bf HMS}({\cal M}')$, then $\Si,{\cal E}\in{\bf
HMS}({\cal M})$ implies $\Si,{\cal E}\insim{\bf HMS}({\cal M}')$, and thus, following
Theorem 1 in \cite{axiomatics},
$\Si,{\cal E}\in{\bf HMS}({\cal M})$ implies $\Delta{\bf M}(\Si,{\cal
E})\leq {\cal M}'$. Now we'll prove that there exists $\Si,{\cal E}\in{\bf HMS}({\cal
M})$ such that ${\cal M}=\Delta{\bf M}(\Si,{\cal
E})$.  Let $\mu_\La$ be such that ${\cal M}_\mu={\cal M}$. We'll define a measure system
$\Si,{\cal E}\in{\bf HMS}(\mu_\La)$. For all $e\in{\cal E}$, let $O_e=\La$, ${\cal
B}_e={\cal B}_\La$, and define the set of strictly classical observables
$\{\varphi_{\la,e}|e\in{\cal E},\la\in\La\}$ such that $\forall p\in\Si,e\in{\cal
E},\forall\la\in\La:\varphi_{\la,e}(p)=\la$. Thus,  $\forall p\in\Si,e\in{\cal E},\forall
B\in{\cal B}_\La:\DL_{p,e}^B=B$, and thus, as a consequence of 
Proposition 2 in \cite{axiomatics}, $\forall p\in\Si,e\in{\cal E},\forall
B\in{\cal B}_e: P_{p,e}(B)=\mu_\La(\DL_{p,e}^B)=\mu_\La(B)$.  Since $\forall p\in\Si,e\in{\cal
E}:P_{p,e}=\mu_\La$, we have $\forall p\in\Si,e\in{\cal E}:{\cal M}_{p,e}={\cal M}_\mu$,
and thus $\Delta{\bf M}(\Si,{\cal E})=\{{\cal M}\}$. 
All this leads us to ${\cal M}\leq {\cal M}'$.
\epf

\section{H.m.s$.$-representations for a given m.s.}

In the previous subsection, we have 'projected' the partial
order structure of ${\bf M}$ on ${\bf M}_{\bf HMS}$.
This allows us to translate the explicit structure of ${\bf M}$ in terms of 'representation
of a m.s$.$ as h.m.s.'$\ $ In fact, the proof of the theorem on the existence of
h.m.s.-representations (see \cite{axiomatics}) was a first example of this approach. In this
subsection, we'll translate some more structural properties of
${\bf M}$ to ${\bf M}_{\bf HMS}$. 
Let us introduce the following
notations referring to the collection of all h.m.s.-representations in ${\bf
HMS}({\bf MX})$, for a given $\Si,{\cal E}\in{\bf MS}$:  
\bit
\item ${\bf HMS}(\Si,{\cal E})=\{\Si',{\cal E}'\in{\bf HMS}({\bf MX})|\Si,{\cal E}\sim\Si',{\cal
E}'\}$ 
\item ${\bf M}_{\bf HMS}(\Si,{\cal E})=\bigl\{{\bf HMS}({\cal M})\in {\bf M}_{\bf HMS}|\Si,{\cal
E}\insim{\bf HMS}({\cal M})\bigr\}$
\eit
One can always enlarge the set of possible h.m.s.-representations if one knows the 'small' 
${\bf HMS}({\cal M})\in{\bf M}_{\bf HMS}(\Si,{\cal E})$, where we consider a h.m.s$.$-representation
smaller than an other one if their respective measure spaces related to $({\cal
B}_\Lambda,\mu_\Lambda)$ have this same ordering (in the sense of Definition 8 in \cite{axiomatics}).
This fact is a straightforward consequence of Theorem 1 in \cite{axiomatics}, and in particular of
the presence of an order condition in the righthandside of eq.15.  

\subsection{General considerations.}

There exists a smallest collection
of mathematical equivalent representations for
$\Si,{\cal E}$ (i.e$.$, a smallest element for the relation $\subsim$ in the collection ${\bf
M}_{\bf HMS}(\Si,{\cal E})$) if there exists ${\cal M}\in {\bf M}$ such that:  
\bit 
\item
$\Si,{\cal E}\insim{\bf HMS}({\cal M})$ 
\item $\forall{\cal M}'\in {\bf M}$ with
$\Si,{\cal E}\insim{\bf HMS}({\cal M}')$: ${\bf HMS}({\cal M})\subsim{\bf
HMS}({\cal M}')$  
\eit 
In the following theorem we prove that in general, there
exists no smallest h.m.s.-representation. First we prove a lemma.    
\blm\label{lm:DLMSiE=NsubM} 
For all ${\bf N}\subseteq {\bf M}$, there exists $\Si,{\cal E}\in
{\bf MS}$ such that: 
\beq
\Delta{\bf M}(\Si,{\cal E})={\bf N}
\eeq
\elm
\bpf
Clearly, for all ${\cal M}\in{\bf N}$, there exists $\mu_{\cal M}\in{\bf MX}$ such that the related
measure space is in ${\cal M}$. Take
${\bf N}$ as set of states and ${\cal E}$ as set of measurements, with for all $e\in{\cal E}$,
${\cal B}_e={\cal B}_{{\Bbb R}}$. For all ${\cal M}\in{\bf N}$ and all $e\in{\cal E}$,
define the outcome probability $P_{{\cal M},e}:{\cal B}_{{\Bbb R}}\rightarrow [0,1]$ such
that $P_{{\cal M},e}=\mu_{\cal M}$. Thus, as the class in ${\bf M}$ related to $P_{{\cal
M},e}$ we find ${\cal M}_{{\cal M},e}={\cal M}$, and thus 
$\Delta{\bf M}({\bf N},{\cal E})={\bf N}$. 
\epf   
We remark that this
proof can also be applied to show that for all ${\bf N}\subseteq {\bf M}$, there exists
$\Si,e\in {\bf MS}$ such that $\Delta{\bf M}(\Si,e)={\bf N}$, i.e$.$, it suffices to
consider one measurement systems to obtain all ${\bf N}\subseteq {\bf M}$.  
\blm\label{th:MnosemiL} 
${\bf M}, \leq$ is not a semi-lattice.  
\elm 
\bpf
We have to give a counterexample, i.e., a set ${\bf N}\subset {\bf M}$ which has no
smallest upper bound. Consider the set
${\bf N}=\{{\cal M}_2^{m_1},{\cal M}_2^{m_2}\}$ where $m_1(1)={2\over 3}$,
$m_1(2)={1\over 3}$, 
$m_2(1)={3\over 4}$ and $m_2(2)={1\over 4}$. We have that both ${\cal M}_3^{m_3}$ and
${\cal M}_3^{m_4}$, defined by  $m_3(1)={2\over 3}$, $m_3(2)={1\over 4}$, $m_3(3)={1\over
12}$, $m_4(1)={1\over 3}$,  $m_4(2)={5\over 12}$ and $m_4(3)={1\over 4}$, are upper
bounds of ${\bf N}$,  but one easily verifies
that ${\cal M}_3^{m_3}\not\leq{\cal M}_3^{m_4}$ and 
${\cal M}_3^{m_4}\not\leq {\cal M}_3^{m_3}$, and thus they cannot be smallest upper
bounds. If
${\cal M}$ would be a smallest upper bound then we should have ${\cal M}_2^{m_1}\leq{\cal
M}$, 
${\cal M}_2^{m_2}\leq{\cal
M}$, ${\cal M}\leq{\cal M}_3^{m_3}$ and 
${\cal M}\leq {\cal M}_3^{m_4}$. But, we also have 
${\cal M}_2^{m_1}\not={\cal M}$, 
${\cal M}_2^{m_2}\not={\cal
M}$, ${\cal M}\not={\cal M}_3^{m_3}$ and 
${\cal M}\not={\cal M}_3^{m_4}$. Due to Lemma 1 in \cite{axiomatics} this is
not possible since there does not exist an element of ${\bf M}$ that fulfills these
conditions.
\epf
\bth\label{th:HMSnoleastel}  
There exist $\Si,{\cal
E}\in {\bf MS}$ such that ${\bf M}_{\bf HMS}(\Si,{\cal E})$ contains no smallest
element for the relation $\subsim$, i.e$.$, the collection ${\bf HMS}(\Si,{\cal E})$ contains no
'smallest' h.m.s.-representation. 
\eth
\bpf  
Following Lemma \ref{th:MnosemiL}, ${\bf M}$ is not a semi-lattice, and as a consequence,
there exists ${\bf N}\subset{\bf M}$ which contains no smallest element. By Lemma
\ref{lm:DLMSiE=NsubM} we know that $\forall {\bf N}\subseteq{\bf M}$ there exists
$\Si,{\cal E}\in {\bf MS}$ such that $\Delta{\bf M}(\Si,{\cal E})={\bf N}$. Thus, there
exists no smallest ${\cal M}\in{\bf M}$ such that $\Delta{\bf M}(\Si,{\cal E})\leq{\cal
M}$, i.e$.$, there exists no
smallest ${\cal M}\in{\bf M}$ such that $\Si,{\cal E}\insim{\bf HMS}({\cal M})$ (see Theorem
1 in \cite{axiomatics}). Thus, as a consequence of Theorem \ref{th:subifleqiflsim}, there
exists no smallest ${\bf HMS}({\cal M})\in {\bf M}_{\bf HMS}$ such that $\Si,{\cal
E}\insim{\bf HMS}({\cal M})$. 
If there
exist a smallest $\Si',{\cal E}'\in{\bf HMS}(\Si,{\cal E})$,
then there exists ${\bf HMS}({\cal M}')\in{\bf M}_{\bf HMS}(\Si,{\cal E})$ with 
$\Si',{\cal E}'\in{\bf HMS}({\cal M}')$. By Theorem \ref{th:subifleqiflsim} we know
that ${\bf HMS}({\cal M}')\subsim{\bf HMS}({\cal M}'')$ for all ${\bf HMS}({\cal
M}'')\in{\bf M}_{\bf HMS}(\Si,{\cal E})$. This contradicts with the first part of this
proof.
\epf 
As a consequence of this theorem, we cannot make general statements concerning a smallest
h.m.s.-representation for a general m.s. Thus, we have to look for
specific m.s$.$ that are such that we can make some explicit statements
concerning the existence (or nonexistence) of a smallest h.m.s.-representation.     
In the next section we will identify the collection of all h.m.s.-representations of the m.s$.$ that
are encountered in quantum mechanics. 

\subsection{H.m.s-representations for quantum-like m.s.}  

In this subsection, we will study the possible h.m.s.-representations for
some  specific classes of m.s$.$ that allow some explicit statements
concerning the existence (or nonexistence) of a smallest h.m.s.-representation. We
will show that these classes of m.s$.$ about which we can make some statements
are of major importance for the case of quantum mechanics, since they contain all quantum
m.s.  In fact, the statements that we are going to prove in this section suffices to
identify all possible h.m.s.-representations for all quantum m.s. 
We have the following expressions that characterize certain quantum-like m.s$.$:
\bit
\item If we consider an entity (e.g$.$ a quantum entity) with only measurements with $n$ outcomes, we
clearly have
$\Delta{\bf M}(\Si,{\cal E})\subseteq{\bf M}_n$. 
\item If we consider a quantum entity with measurements with a finite
number of outcomes we have ${\bf M}_{{\Bbb X}}=\Delta{\bf M}(\Si,{\cal E})$, as a consequence of
the definition of a quantum state.
\item If we consider a quantum entity with measurements with an at most
countable number of outcomes we have ${\bf M}_{{\Bbb N}}=\Delta{\bf M}(\Si,{\cal E})$.
\eit
As we will show at the end of this section, these expressions characterize the quantum m.s$.$ in a
sufficient way to identify all their h.m.s$.$-representations. We start with a theorem which states
that if
${\bf M}_{{\Bbb N}}\subseteq\Delta{\bf M}(\Si,{\cal E})$, then this m.s$.$ has no
smaller h.m.s.-representation than the ones contained in ${\bf HMS}([0,1])$,
i.e$.$, there exists a smallest h.m.s.-representation, but it is one that needs
the 'maximal' number of strictly classical observables, namely a continuous set.
\blm\label{finhassub} 
${\bf M}_{{\Bbb X}}$ has a smallest upper bound in ${\bf
M}$, namely  ${\cal M}_{{\Bbb R}}$.
\elm
\bpf
We only have to prove that there exists no smaller upper bound 
for ${\bf M}_{{\Bbb X}}$ in ${\bf
M}$ than ${\cal M}_{{\Bbb R}}$, the greatest element of ${\bf M}$ itself. Suppose
that ${\cal M}$ is such a smaller upper bound. If ${\cal M}\not= {\cal M}_{{\Bbb
R}}$ and ${\cal B},\mu\in {\cal M}$, there exists  a set $B\in {\cal B}$ such that for
all $B'\in {\cal B}$, $B'\subseteq B$ implies $B'=B$ or $B'=\emptyset$ (see Lemma 2 in
\cite{axiomatics}). Let $\mu(B)=a$. Define $N$ as the smallest integer such that
$N>1/a$. Consider ${\cal M}_N^m$ where $\forall i\in{\Bbb X}_N:m(i)=1/N$. Since ${\bf
M}_{{\Bbb X}}\leq {\cal M}$ we have ${\cal M}_N^m\leq {\cal M}$. Thus, there exists
a $\si$-morphism  $F:{\cal B}_{{\Bbb
N}}\rightarrow {\cal B}$ fulfilling Definition 8 in \cite{axiomatics}.  Clearly, for all
$i\in{\Bbb X}_N:B\cap F(\{i\})=\emptyset$ or $B\cap F(\{i\})=B$.
Since for all $i\in{\Bbb
X}_N$ we have $F(\{i\})\in {\cal B}$ and $\cup_{i\in{\Bbb N}}(B\cap F(\{i\}))=B\cap 
(\cup_{i\in{\Bbb N}}F(\{i\}))=B\cap I=B$, there exists $i\in{\Bbb X}_N$ such 
that $B\cap
F(\{i\})=B$, i.e., $B\subseteq F(\{i\})$, and thus, $\mu(F(\{i\}))\geq a$. This contradicts with
$\mu(F(\{i\}))=m(i)=1/N<a$.
\epf   
\blm\label{th:MN,smallupb.} 
${\bf M}_{{\Bbb N}}$ has a smallest upper bound in ${\bf M}$, namely 
${\cal M}_{{\Bbb R}}$.
\elm
\bpf
If we consider ${\cal M}_{{\Bbb N}}^m$, where for all $i$ in ${\Bbb
X}_{N-1}: m(i)=1/N$, and for all integers $i\geq N: m(i)=N^{-1}.2^{N-i-1}$ ($m$ defines
a probability measure because 
$\sum_{i\in{\Bbb N}}m(i)=(N-1)(1/N)+\sum_{i\in{\Bbb N}}(2^{-i}/N)
=(N-1+\sum_{i\in{\Bbb N}}2^{-i})/N=1$), we can
prove this proposition along the same lines as the proof of Lemma \ref{finhassub}.   
\epf
\bth\label{MNsubeqDMSE}  
Let $\Si,{\cal
E}\in{\bf MS}$. We have:
\beq
{\bf M}_{{\Bbb N}}\subseteq\Delta{\bf M}(\Si,{\cal E})
\Longrightarrow{\bf HMS}(\Si,{\cal E})\subsim{\bf HMS}([0,1]) \eeq
\eth
\bpf
Following Theorem 3 in \cite{axiomatics} we know that 
${\bf HMS}([0,1])={\bf HMS}({\cal M}_{{\Bbb R}})\in{\bf M}_{\bf
HMS}(\Si,{\cal E})$. If there exists ${\cal M}\in{\bf M}$ such that ${\cal M}<{\cal
M}_{{\Bbb R}}$ and ${\bf HMS}({\cal M})\in{\bf M}_{\bf HMS}(\Si,{\cal E})$, then,
as a consequence of Theorem 1 in \cite{axiomatics}, $\Delta{\bf M}(\Si,{\cal E})\leq{\cal
M}$. Thus,
${\bf M}_{{\Bbb N}}\leq{\cal M}$. This contradicts with Lemma \ref{th:MN,smallupb.}
which states that ${\cal M}_{{\Bbb R}}$ is the supremum of ${\bf M}_{{\Bbb N}}$,
and thus, there exist no smaller upper bounds. As a consequence, ${\bf M}_{\bf
HMS}(\Si,{\cal E})= \{{\bf HMS}({\cal M}_{{\Bbb R}})\}$, and ${\bf
HMS}(\Si,{\cal E})\subsim {\bf HMS}([0,1])$. 
\epf
We proceed with a theorem which states that the smallest
h.m.s.-representations in the case that ${\bf M}_{{\Bbb
X}}\subseteq\Delta{\bf M}(\Si,{\cal E})$ is also contained in ${\bf HMS}([0,1])$.  
\bth\label{MXsubeqDMSE}  
Let $\Si,{\cal
E}\in{\bf MS}$. We have:
\beq
{\bf M}_{{\Bbb X}}\subseteq\Delta{\bf M}(\Si,{\cal E})\Longrightarrow
{\bf HMS}(\Si,{\cal E})\subsim{\bf HMS}([0,1]) 
\eeq
\eth
\bpf
The proof of this theorem is the same as the proof of Theorem \ref{MNsubeqDMSE}, if we 
replace 'Lemma \ref{th:MN,smallupb.}' by 'Lemma \ref{finhassub}'.
\epf
In the last two theorems of this section we identify measurement systems that do have
smaller h.m.s.-representations than the ones contained in 
${\bf HMS}([0,1])$, but for which there exists no smallest ones. More precisely, if we consider 
${\bf M}_n$ for some fixed $n\in{\Bbb N}$ in stead of ${\bf M}_{{\Bbb X}}$,
then, measurement systems $\Si, {\cal E}\in {\bf MS}$ fulfilling ${\bf
M}_n=\Delta{\bf M}(\Si,{\cal E})$ do have smaller h.m.s.-representations
than the ones fulfilling ${\bf M}_{{\Bbb X}}=\Delta{\bf M}(\Si,{\cal E})$,
but there doesn't exists a smallest one.  
\blm\label{bindecomp}
Let $a\in [0,1]$. There exist one and at most two sequences $(a_i)_i$, with 
$\forall i\in{\Bbb N}:a_{i}\in\{0,1\}$ and such that:
\beq
a=\sum_{i\in{\Bbb N}}a_i/2^i
\eeq  
i.e$.$, there are at most two non-equal subsets $B_1,B_2\in{\Bbb N}$ such that:
\beq\label{eq2bindecomp}
\sum_{i\in B_1}1/2^i= \sum_{i\in B_2}1/2^i
\eeq 
\elm
\bpf
This lemma expresses the well known representation of reals as
decimal numbers, but now in the scale $2$ in stead of the scale $10$. For a detailed proof
we refer to \cite{har38} p.107-112. Here, we'll only sketch how these two possible
decompositions can be constructed.
A first decomposition is defined by: 
$\forall i\in{\Bbb N}$ such that $a-\sum_{j\in{\Bbb X}_{i-1}}a_j/2^j>1/2^i$: 
$a_i=1$, otherwise: $a_i=0$.  
A second decomposition is defined by:
$\forall i\in{\Bbb N}$ such that $a-\sum_{j\in{\Bbb X}_{i-1}}a_j/2^j\geq 1/2^i$: 
$a_i=1$, otherwise: $a_i=0$.  
One
can prove that for both decompositions, we find the same sum $a=\sum_{i\in{\Bbb
N}}a_i/2^i$, and that for almost\footnote{We mean for all $a\in [0,1]$ except for a set
of measure zero.} all $a\in [0,1]$, both decompositions give the same sequence
$(a_i)_i$. If we define $B_1=\{j|a_j=1\}$ for the first decomposition, and
$B_2=\{j|a_j=1\}$ for the second one, we find eq.(\ref{eq2bindecomp}).   
\epf   
\blm\label{lm:lemcomp}  
If $\forall i\in{\Bbb X}_n$,
$Q_i\in{\Bbb R}$ and $m_i=Q_i.(1-\sum_{j=1}^{i-1}m_j)$ then: 
\beq\label{lemcomp}
\forall
i\in{\Bbb X}_n:m_i=Q_i.\prod_{j=1}^{i-1}(1-Q_j)
\eeq
\beq
1-\sum_{j=1}^{n}m_i=\prod_{i=1}^{n}(1-Q_i)
\eeq
\elm
\bpf
We prove this lemma by induction. For $n=1$ we have
$m_1=Q_1$ and $1-m_1=1-Q_1$. If the lemma is true $\forall n'\in{\Bbb
X}_{n-1}$, then eq.\ref{lemcomp} is true $\forall i\in{\Bbb X}_{n-1}$. 
\beqa
m_n&=&Q_n(1-\sum_{i=1}^{n-1}m_i)=Q_n(1-\sum_{i=1}^{n-1}Q_i\prod_{j=1}^{i-1}(1-Q_j))\cr
&=&Q_n((1-Q_1)-\sum_{i=2}^{n-1}Q_i\prod_{j=1}^{i-1}(1-Q_j))\cr
&=&Q_n(1-Q_1)(1-\sum_{i=2}^{n-1}Q_i\prod_{j=2}^{i-1}(1-Q_j))\cr
&=&Q_n\prod_{i=1}^{n-1}(1-Q_i)\cr
1-\sum_{j=1}^{n}m_j
&=&(1-\sum_{i=1}^{n}Q_i\prod_{j=1}^{i-1}(1-Q_j))\cr
&=&(1-Q_1)(1-\sum_{i=2}^{n}Q_i\prod_{j=2}^{i-1}(1-Q_j))\cr
&=&\prod_{i=1}^{n}(1-Q_i)\cr
\eeqa
what completes the proof.
\epf
\blm\label{th:Mnsmalluppb.}
For all $n\in{\Bbb N}$, ${\bf M}_n$ has an upper bound in 
${\bf M}_{{\Bbb N}}$. 
\elm
\bpf
For all $m\in M_n$,
let $Q_1=m(1)$, and $\forall i\in{\Bbb X}_{n-1}\setminus\{1\}$: 
let $Q_i=0$ if $1-\sum_{j=1}^{i-1}m(j)=0$
and let $Q_i=m(i)(1-\sum_{j=1}^{i-1}m(j))^{-1}$ if 
$1-\sum_{j=1}^{i-1}m(j))\not=0$. Since for all $i\in{\Bbb
X}_{n-1}$, $Q_i\in[0,1]$, we can define a sequence $(q_{i,j})_j$, with 
$\forall i,j\in{\Bbb N}:q_{i,j}\in\{0,1\}$, such
that $(q_{i,j})_j$ represents a binary decomposition\footnote{See Lemma
\ref{bindecomp}.} of $Q_i$ fulfilling  $\forall i\in{\Bbb X}_{n-1}
:Q_i=\sum_{j\in{\Bbb N}}(q_{i,j}/2^j)$. Since ${\Bbb N}^{n-1}$ is countable,
there exists a one-to-one map $\xi:{\Bbb N}\rightarrow{\Bbb N}^{n-1}$. We
choose one fixed $\xi$, and $\forall i\in{\Bbb N}$ we denote $\xi
(i)\in{\Bbb N}^{n-1}$ as $i_1,i_2,\ldots,i_{n-1}$. 
Define ${m'}\in M_{{\Bbb N}}$ such that
$\forall i\in{\Bbb N}:m'(\{i\})=\prod_{j=1}^{m-1}(1/2^{i_j})$. $m'$
defines a probability measure (all summations are taken over ${\Bbb N}$):
\beqa
m'({\Bbb N}) &=&\sum_i\prod_{j=1}^{m-1}(1/ 2^{i_j})
=\sum_{i_1}\sum_{i_2}\ldots\sum_{i_{n-1}}\prod_{j=1}^{n-1}(1/ 2^{i_j})\\
&=&\sum_{i_1}(1/
2^{i_1})\sum_{i_2}(1/ 2^{i_2})\ldots\sum_{i_{n-1}} (1/ 2^{i_{n-1}}) =\bigl(\sum_{j}
1/2^j\bigr)^{n-1}=1
\eeqa
We define $F:{\cal B}_n\rightarrow {\cal B}_{{\Bbb
N}}$ such that for all $j\in{\Bbb X}_{n-1}$:
\beqa
F(\{n\})&=&\{i|i\in{\Bbb N},\forall j\in{\Bbb X}_{n-1}:q_{j,i_j}=0\}\\  
F(\{j\})&=&\{i|i\in{\Bbb N},\forall k\in{\Bbb X}_{j-1}:q_{k,i_k}=0, q_{j,i_j}=1\}
\eeqa 
and for all $B\in {\cal B}_n:
F(B)=\cup_{i\in B}F(\{i\})$. For all $i,j\in{\Bbb X}_n$ such that $i\not=j$
(i.e$.$, $\{i\}\cap \{j\}=\emptyset$) we have $F(\{i\})\cap F(\{j\})=\emptyset$.
Since: 
\beqa
F({\Bbb X}_n)&=&\cup_{i\in{\Bbb N}}F(\{i\})\\
&=&\{i|j\in{\Bbb X}_{n-1},q_{j,i_j}=1,\forall k\in{\Bbb
X}_{j-1}:q_{k,i_k}=0\}\\
& &\cup\{i|\forall j\in{\Bbb X}_{n-1}:q_{j,i_j}=0\}\\
&=&\{i| j\in{\Bbb X}_{n-1}:q_{j,i_j}=1\ {\rm or}\ \forall j\in{\Bbb
X}_{n-1}:q_{j,i_j}=0\}={\Bbb N}
\eeqa
$F$ is a $\si$-morphism. 
We still have to verify if $\mu_m=\mu_{m'}$ (see Definition 8 in \cite{axiomatics}). For all
$j\in{\Bbb X}_{n-1}$: 
\beqa 
m'(j)&=&\sum_{i\in F(j)}\prod_{k=1}^{m-1}{1\over 
2^{i_k}}=\sum_{i\in F(j)}(\prod_{k=1}^{j-1}(1-{q_{k,i_k}\over 2^{i_k}})){q_{j,i_j}\over
2^{i_j}}\prod_{k=j+1}^{m-1}{1\over 2^{i_k}}\cr
&=&\sum_{i\in{\Bbb
N}}(\prod_{k=1}^{j-1}(1-{q_{k,i_k}\over 2^{i_k}})){q_{j,i_j}\over
2^{i_j}}\prod_{k=j+1}^{m-1}{1\over
2^{i_k}}\cr
&=&\sum_{i_1\in{\Bbb N}}\sum_{i_2\in{\Bbb N}}\ldots\sum_{i_{m-1}\in{\Bbb
N}}(\prod_{k=1}^{j-1}(1-{q_{k,i_k}\over 2^{i_k}})){q_{j,i_j}\over
2^{i_j}}\prod_{k=j+1}^{m-1}{1\over
2^{i_k}}\cr
&=&\prod_{k=1}^{j-1}(\sum_{{i_k}\in{\Bbb
N}}(1-{q_{k,i_k}\over 2^{i_k}}))\sum_{{i_j}\in{\Bbb
N}}{q_{j,i_j}\over 2^{i_j}} 
\prod_{k=j+1}^{m-1}\sum_{{i_k}\in{\Bbb N}}{1\over
2^{i_k}}\cr
&=&\prod_{k=1}^{j-1}(\sum_{k\in{\Bbb N}}{1\over
2^k}-\sum_{k\in{\Bbb N}}{q_{k,k}\over 2^k})\sum_{k\in{\Bbb
N}}{q_{j,k}\over 2^k}\prod_{k=j+1}^{m-1}\sum_{k\in{\Bbb N}}{1\over
2^k}\cr
&=&\prod_{k=1}^{j-1}(1-\sum_{k\in{\Bbb N}}{q_{k,k}\over 2^k}\sum_{j\in{\Bbb
N}}{q_{j,j}\over 2^j})\cr 
&=&\prod_{k=1}^{j-1}(1-Q_k) Q_j= m(j)\cr
m'(n)&=&\sum_{i\in
F(n)}\prod_{j=1}^{m-1}(1-{q_{j,i_j}\over
2^{i_j}})=\prod_{j=1}^{m-1}(1-Q_j)=1-\sum_{j=1}^{m-1}m(j)=m(n)   
\eeqa
as a consequence of Lemma \ref{lm:lemcomp}. 
\epf
\bth\label{MnsubeqDMSE} 
Let $\Si,{\cal E}\in{\bf MS}$ and $n\in{\Bbb N}$. We have:
\beq
\Delta{\bf M}(\Si,{\cal E})\subseteq {\bf M}_n\Longrightarrow
\Si,{\cal E}\insim {\bf HMS}(\Si,O_{\cal E},{\Bbb N})
\eeq
\eth
\bpf
For all $n\in{\Bbb N}$ we know that ${\bf M}_n$ has an upper bound ${\cal M}\in{\bf
M}_{{\Bbb N}}$ (see Lemma \ref{th:Mnsmalluppb.}). 
Since ${\cal M}\in{\bf
M}_{{\Bbb N}}$ we have ${\bf HMS}({\cal M})\subsim {\bf HMS}({\Bbb N})$. Thus,
if ${\bf M}_n=\Delta{\bf M}(\Si,{\cal E})$, then $\Delta{\bf M}(\Si,{\cal E})\leq{\cal M}$
and thus, according to Theorem 1 in \cite{axiomatics},  we have $\Si,{\cal E}\insim{\bf
HMS}({\cal M})\subsim {\bf HMS}({\Bbb N})$. Thus, $\Si,{\cal E}\insim {\bf
HMS}(\Si,O_{\cal E},{\Bbb N})$.
\epf
\blm\label{th:Mnsmalluppbis.}
For all $n\in{\Bbb N}$, there exists no smallest upper bound for ${\bf M}_n$ in
${\bf M}$.  
\elm
\bpf
We prove that there exists no smallest upper bound in ${\bf M}_{{\Bbb N}}$ for the case
$n=2$. For $n>2$, the proof goes along the same lines, but gets very complicated on a
notational level. From the proof of Lemma \ref{th:Mnsmalluppb.} we know that ${\cal
M}_{{\Bbb N}}^m$, with
$\forall i\in{\Bbb N}: m(\{i\})=1/2^i$, is an upper bound. 
First we will prove that there exists no upper bound for
${\bf M}_2$, smaller than ${\cal M}_{{\Bbb N}}^m$. Suppose that 
${{\cal M}_{{\Bbb N}}^m}'$
is such a smaller upper bound (it's clear  that such a smaller upper bound cannot be an
element of ${\bf M}_{n'}$ for some $n'\in{\Bbb N})$. Let  $F:{\cal B}_{{\Bbb
N}}\rightarrow {\cal B}_{{\Bbb N}}$ be a $\si$-morphism fulfilling Definition 8 in
\cite{axiomatics}, and which is not onto, i.e., there exists $i\in {\Bbb N}$ with
$\{i\}\not\subset
\{F(B)|B\in {\cal B}_{{\Bbb N}}\}$ (otherwise,
$\{F(B)|B\in {\cal B}_{{\Bbb N}}\}={\cal B}_{{\Bbb N}}$ because of the $\si$-additivity). Since $i\in{\Bbb
N}=\cup_{j\in{\Bbb N}}F(\{j\})$, there exists $j\in{\Bbb N}$ such that $i\in
F(\{j\})$. Denote $B=F(\{j\})$, and let $n\in B$ be such that there exists $n'\in
B:n'<n$.  Let $B_1=\{n, n+1\}$ and
$B_2=\{n\}\cup \{i|i\in{\Bbb N},i>n+1\}$. Since
$1/2^n+1/2^{n+1}=1/2^n+\sum_{i=n+1}^{i=\infty}1/2^{i+1}$, we have $\sum_{i\in B_1}1/2^i=
\sum_{i\in B_2}1/2^i$. For all $i\in B_1$ and for all $i\in B_2$ we have $i\geq n> n'$, and
thus we have $B\cap B_1\subset B$ and  $B\cap B_2\subset B$. Thus, $B_1\not\in
\{F(B)|B\in {\cal B}_{{\Bbb N}}\}$ and $B_2\not\in \{F(B)|B\in {\cal B}_{{\Bbb
N}}\}$. Consider ${\cal M}_2^{m''}$ such that $m''(1)=1/2^n+1/2^{n+1}$. 
Since ${{\cal
M}_{{\Bbb N}}^m}'$ is an upper bound for ${\bf M}_2$, there exists a
$\si$-morphism  $F':{\cal B}_2\rightarrow {\cal B}_{{\Bbb N}}$
fulfilling Definition 8 in \cite{axiomatics}.
Let $B_3=F\circ F'(\{1\})$. We have
$\mu_{m}(B_3)=
\mu_{m''}(\{1\})=1/2^n+1/2^{n+1}$, and thus, $\sum_{i\in B_3}1/2^i=1/2^n+1/2^{n+1}$. Since 
$B_3\in\{F(B)|B\in {\cal B}_{{\Bbb N}}\}$ we have $B_1\not=B_3$ and $B_2\not=B_3$, and
thus,  there are three subsets of ${\Bbb N}$ such that $\sum_{i\in B_1}1/2^i= 
\sum_{i\in B_2}1/2^i=\sum_{i\in B_3}1/2^i$. This contradicts with Lemma
\ref{bindecomp}.
Now we'll construct an upper bound ${\cal M}_{{\Bbb N}}^{m'}$ such that 
${\cal M}_{{\Bbb N}}^{m'}\not\geq{\cal M}_{{\Bbb N}}^{m}$. Consider ${\cal
M}_{{\Bbb N}}^{m'}$ where $m'(1)=m'(2)=1/3$, and for all
integers $i\geq 3: m'(i)=(3.2^{i-2})^{-1}$. If 
${\cal M}_{{\Bbb N}}^{m'}\geq{\cal M}_{{\Bbb N}}^{m}$, there exists a 
$\si$-morphism $F:{\cal B}_{{\Bbb N}}\rightarrow {\cal B}_{{\Bbb N}}$ fulfilling
Definition 8 in \cite{axiomatics}. Since,
$\cup_{i\in{\Bbb N}}F(\{i\})={\Bbb N}$, there exist $i,j\in{\Bbb N}$ such
that $1\in F(\{i\})$ and
$2\in F(\{j\})$.  If $i=j$ then $\{1,2\}\subseteq F(\{i\})$ and thus, $\mu_m(\{i\})\geq
m'(1)+m'(2)=2/3$, which is not possible.   Thus we
have  $m(i)=\mu_{m'}(F(\{i\}))\geq \mu_{m'}(\{1\})=m'(1)=1/3$ and
$m(j)=\mu_{m'}(F(\{j\}))\geq\mu_{m'}(\{2\})=m'(2)=1/3$, and this contradicts with
$\{i|i\in{\Bbb N},m(i)\geq 1/3\}=\{1\}$. We still have to prove that 
${\cal M}_{{\Bbb N}}^{m'}\geq {\bf M}_2$. 
Consider ${\cal M}_2^{m''}$ such that $m''(1)=a$ and $m''(2)=1-a$. If $a<1/3$, let
$B_1=\emptyset$ and $a'=3a$. If $1/3 \leq a<2/3$, let
$B_1=\{1\}$ and $a'=3a-1$. If $a\geq 2/3$ , let
$B_1=\{1,2\}$ and $a'=3a-2$. Consider a binary decomposition 
$a'=\sum_{i\in{\Bbb N}}a'_i/2^i$ and define $B_2=\{i|a'_i=1\}$. 
One easily verifies that the map $F:{\cal B}_2\rightarrow {\cal B}_{{\Bbb N}}$
defined by $F(\{1\})=B_1\cup B_2$ and $F(\{2\})={\Bbb N}\setminus (B_1\cup B_2)$ fulfills
Definition 8 in \cite{axiomatics}. 
\epf
\bth\label{MnsubeqDMSEbis} 
$\forall\Si,{\cal E}\in{\bf MS}$, $n\in{\Bbb N}$ with ${\bf M}_n=\Delta{\bf
M}(\Si,{\cal E})$:
${\bf HMS}(\Si,{\cal E})$ contains no smallest h.m.s.-representation.
\eth
\bpf
If there exists ${\bf HMS}({\cal M})\in{\bf M}_{\bf HMS}(\Si,{\cal E})$ such that for all 
${\bf HMS}({\cal M}')\in{\bf M}_{\bf HMS}(\Si,{\cal E})$ we have
${\bf HMS}({\cal M})\leq{\bf HMS}({\cal M}')$. Then, as a consequence of Theorem 1 in
\cite{axiomatics}, ${\bf HMS}({\cal M})$ is the smallest element in ${\bf M}_{\bf HMS}$ such
that $\Delta{\bf M}(\Si,{\cal E})\leq{\cal M}$, i.e$.$, 
${\cal M}$ is the smallest element in ${\bf M}$
such that 
${\bf M}_n\leq{\cal M}$ (see Theorem \ref{th:subifleqiflsim}). This contradicts with Lemma
\ref{th:Mnsmalluppbis.} which states that ${\bf M}_n$ has no smallest upper bound in ${\cal M}$.
\epf
With these theorems we are now able to identify the possible h.m.s$.$-representations of quantum m.s.
The different possible candidates for a h.m.s.-representation correspond with the different
possible measure spaces in ${\bf M}$, and which are given by Lemma 1 in \cite{axiomatics}, and
which are essentially measure spaces related to a continuous, a countable or a finite set of states
of the measurement context. From Theorem
\ref{MNsubeqDMSE} and Theorem
\ref{MXsubeqDMSE} follows that if we consider quantum m.s$.$ with measurements with an unbounded
number of outcomes, we are not able to find smaller h.m.s.-representations than the one we have
identified in Theorem 3 in
\cite{axiomatics}, i.e., a h.m.s.-representation with a continuous set of states of the measurement
context.  If we consider a m.s$.$ with measurements with $n$-outcomes (e.g$.$ a
spin-${{n-1}\over 2}$ quantum entity) we find h.m.s$.$-representations with a countable set of
states of the measurement context.  Nonetheless, there exists no 'preferred' smallest
h.m.s$.$-representation\footnote{This obviously also implies that there exist no 
h.m.s$.$-representations with a finite set of
states of the measurement context.}. 

\section{Two explicit examples.}\label{examples}

We end this paper with two examples, namely Aerts' model system for a spin-${1\over 2}$
quantum entity (introduced in \cite{aer86} as an abstraction of the model in \cite{gispir}, discussed
in more detail in for example
\cite{thesisbob} and
\cite{cza92}), and a 'reduced' version of this model system that is implemented by the
results of this paper.  These two model systems will enable us to visualize the results of this
paper.  We start with Aerts' model system\footnote{Since Aerts' model system has already been
published many times, we won't go in to much details. For these details we refer to some of these
earlier publications by Aerts and his collaborators.}.
As a representation of the states of a spin-${1\over 2}$ quantum entity we consider the Poincar\'e
representation, i.e., all states are represented on a sphere in ${\Bbb R}^3$, orthogonal states
correspond with antipodic point on the sphere. If a point has 
coordinates $v$,
we denote the state corresponding to this point as $p_v$. A measurement $e_u$ on the 
entity
in a state $p_v$ is defined in the following way:  
\bit 
\item Consider a straight line segment with one of its endpoints in the point $u$ of the
sphere, and the other endpoint in the diametrically opposite point $-u$. We'll denote this
segment as $[u,-u]$.
\item We project $v$ orthogonally on $[-u,u]$ and obtain the point $v'$.  This point
defines two segments $[-u,v']$ and  $[v',u]$ (see Fig. 1).
\item Consider a stochastic variable $\lambda$ located on the segment $[-u,u]$, and
suppose that relative frequency of appearance of the possible $\lambda$
is uniformly distributed on $[-u,u]$.  If
$\lambda\in[-u,v']$, the point corresponding to the state of the entity  moves to $u$
along $[v',u]$ and we obtain a state $p_u$.  If $\lambda\in]v',u]$, the point moves to
$-u$ along $[v',-u]$ and we obtain $p_{-u}$.      
\eit
\noindent As a consequence, there are two outcome states for this measurement $e_u$: $p_u$
and $p_{-u}$.  As been shown in earlier publications, this model system gives rise to the same
probability structure as a spin-${1\over 2}$ quantum entity. 

\vskip -0.1 cm\par\noindent 
\hskip 4.4 cm {\hbox{\vrule height 4 truecm width 0pt
\epsfig{file=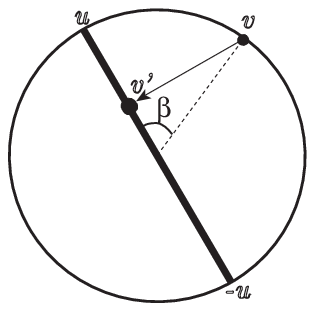}\vrule height 0pt width 9truecm}} 

\bigskip\noindent Fig.1: Illustration of a measurement $e_u$ on an Aerts' spin-${1\over 2}$ entity
when the initial state is $p_v$.

\bigskip\noindent Within the formalism of \cite{axiomatics} and this paper, $\Si$ corresponds in a
one-to-one way with the sphere $S$,
$\La$ corresponds with the line segment
$[u,-u]$, 
$\mu_\La$ is the uniformly distributed probability measure of the stochastic variable
$\la$ and $O_e=\{p_{u}, p_{-u}\}$. 
For a clear description of $\varphi_\la$ we refer to Fig. 2.  The different measurements in
${\cal E}$ correspond with the different possible choices of antipodic points, and their
h.m.s$.$-representations are related in the way as described in section 3.3 in
\cite{axiomatics}. Since it is shown in Theorem
\ref{MnsubeqDMSE} that for measurements with a finite and bounded number of outcomes there exists a
h.m.s.-representation with
$\Lambda={\Bbb N}$, we should be able to find such a 'reduced' representation for
Aerts' spin-${1\over 2}$ quantum entity. The straightforward way to do this is by
explicitly applying the model used in the proof of Lemma
\ref{th:Mnsmalluppb.}\footnote{For the specific case of $n=2$, the rather complicated
model of Lemma \ref{th:Mnsmalluppb.} reduces to a rather simple model system.}.  We find
the model system as outlined in Fig. 2. 
 
\section{Conclusion.}     

We were able to identify the possible h.m.s$.$-representations of quantum m.s., and we showed that
it suffices to know the 'small' h.m.s$.$-representations in order to know the complete collection of
all h.m.s$.$-representations. Quantum m.s$.$ with measurements with an unbounded number of outcomes
(i.e., a finite but unbounded, an infinite but countable, or a continuous number of outcomes)  all
have a 'preferred' smallest h.m.s.-representation with a continuous set of states of the measurement
context. Quantum m.s$.$ with a finite and bounded number of outcomes (e.g. spin-${{n-1}\over 2}$
quantum m.s$.$) have h.m.s.-representation with a countable set of states of the measurement context
(see for example the alternative for Aerts' spin-${1\over 2}$ h.m.s.-representation in section
\ref{examples}), but they do not have a preferred smallest h.m.s.-representation.

\vskip 0.3 cm\par\noindent
\hskip 0.7 cm {\hbox{\vrule height 10.5 truecm width 0pt
\epsfig{file=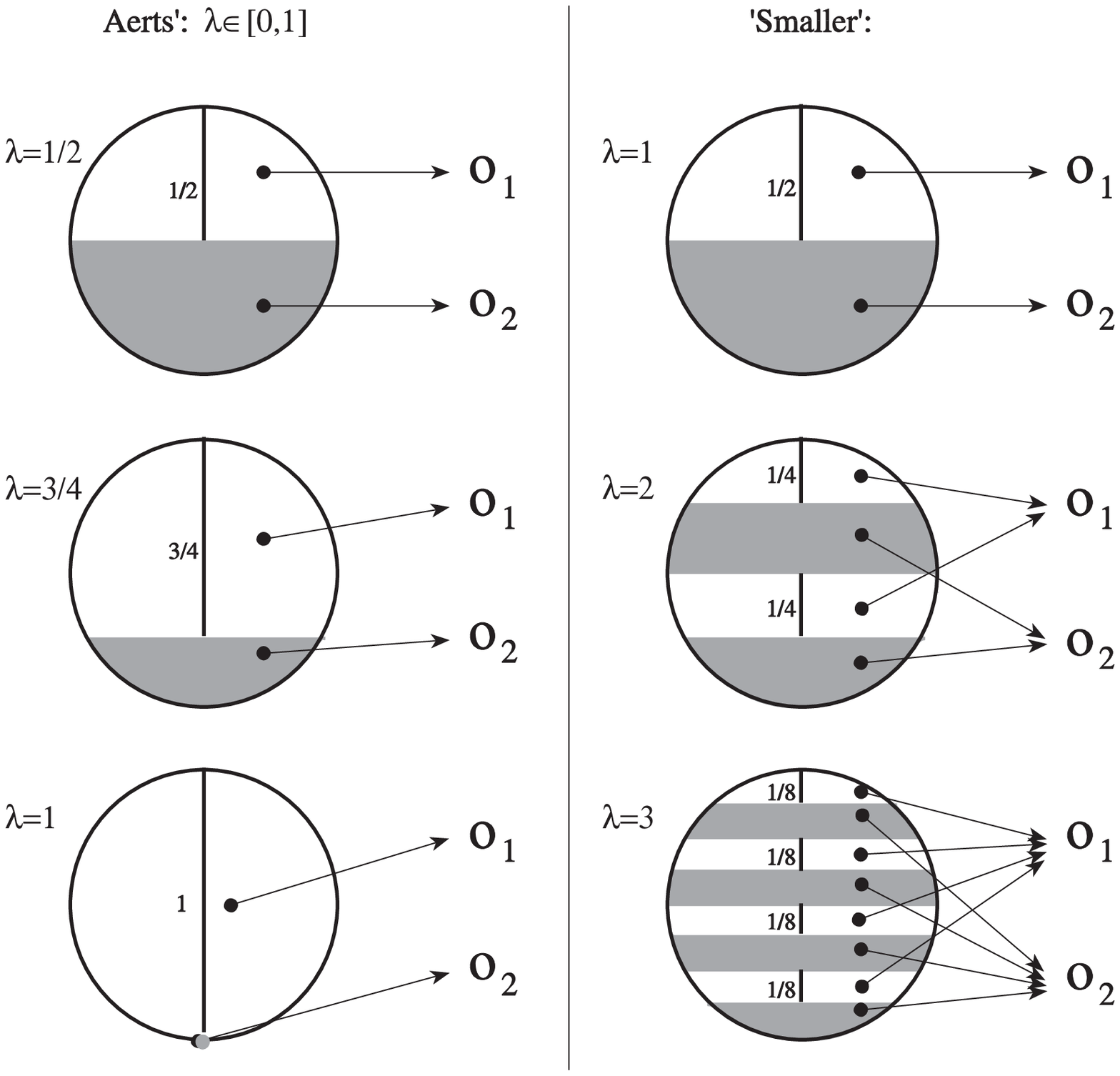,width=.9\textwidth}\vrule height 0pt width  
9truecm}}   

\noindent Fig. 2: A comparison of Aerts' h.m.s$.$ with a 'smaller' h.m.s$.$ for a
spin-${1/2}$ quantum entity. In the case of Aerts' h.m.s$.$, for every $\la\in [0,1]$ the state
space is divided in an upper and a lower half, depending on the value of $\la$. If the entity is in
a state located in the upper half we obtain an outcome $o_1$ and if the entity is in a state located
in the lower half we obtain an outcome $o_2$. The arrows in the drawing correspond with the map
$\varphi_\la:\Si\rightarrow O_e$. In the case of the 'smaller' h.m.s$.$,  for every
$\la\in{\Bbb N}$ the state space is divided in
$2^\la$ spherical bands. If the entity is in a state located in the upper band we obtain an outcome
$o_1$, if it is in a state located in the second band we obtain an outcome $o_2$, if it is in a
state located in the third band we obtain an outcome $o_1$, ... This h.m.s$.$ is completely defined
by the relation $\forall\la\in{{\Bbb N}}$: $\mu_{\Bbb N}(\{\la\})=1/2^\la$. One easily
verifies that this 'smaller' h.m.s$.$ is mathematically equivalent with Aerts' h.m.s$.$
  
\section{Acknowledgments.}     
   
This work is supported by the IUAP-III n$^{\rm o}$9. The author is Research
Assistant of the National Fund for Scientific Research.

\end{document}